# Strategic analysis of hydrogen market dynamics across collaboration models

Mohammad Asghari [a], Hamid Afshari [a,*], Mohamad Y. Jaber [b,1,2,3], Cory Searcy [b]

[a] Department of Industrial Engineering Dalhousie University, Sexton Campus, 108 - 5269 Morris Street, PO Box 15000, Halifax, NS, B3H4R2, Canada
[b] Department of Mechanical, Industrial, and Mechatronics Engineering, Toronto Metropolitan University, 350 Victoria Street, Toronto, ON, M4B 2K3, Canada

## ABSTRACT

The global energy landscape is experiencing a transformative shift, with an increasing emphasis on sustainable and clean energy sources. Hydrogen remains a promising candidate for decarbonization, energy storage, and as an alternative fuel. This study explores the landscape of hydrogen pricing and demand dynamics by evaluating three collaboration scenarios: market-based pricing, cooperative integration, and coordinated decision-making. It incorporates price-sensitive demand, environmentally friendly production methods, and market penetration effects, to provide insights into maximizing market share, profitability, and sustainability within the hydrogen industry. This study contributes to understanding the complexities of collaboration by analyzing those structures and their role in a fast transition to clean hydrogen production by balancing economic viability and environmental goals. The findings reveal that the cooperative integration strategy is the most effective for sustainable growth, increasing green hydrogen's market share to 19.06 % and highlighting the potential for environmentally conscious hydrogen production. They also suggest that the coordinated decision-making approach enhances profitability through collaborative tariff contracts while balancing economic viability and environmental goals. This study also underscores the importance of strategic pricing mechanisms, policy alignment, and the role of hydrogen hubs in achieving sustainable growth in the hydrogen sector. By highlighting the uncertainties and potential barriers, this research offers actionable guidance for policymakers and industry players in shaping a competitive and sustainable energy marketplace.

## 1. Introduction

The hydrogen market landscape is evolving rapidly with a growing demand for clean hydrogen. Almost all industrial hydrogen producers today use the fossil fuel-based steam methane reforming method. The produced hydrogen is known as grey hydrogen. However, the demand for grey hydrogen will decline as the demand for clean hydrogen rises and its cost becomes more competitive [1]. Clean hydrogen production today faces high costs and significant technological challenges that hinder its widespread adoption. By 2050, clean hydrogen demand could account for 73–100 % of the total hydrogen demand [1]. This projected surge in demand underscores the necessity for policymakers, industry players, and investors to adapt to the evolving market landscape to facilitate scalability and make informed decisions regarding production strategies and strategic positioning within the hydrogen market.

Hydrogen is being looked at as a key enabler in the broader energy transition narrative due to its potential role in decoupling economic growth from carbon emissions. Decoupling is crucial to achieving ecological sustainability, as it allows economies to grow without a corresponding increase in environmental degradation. Vadén et al. [2] and Basu et al. [3] explored decoupling strategies concluding how vital they are for reducing dependence on fossil fuels and promoting renewable energy sources. Hydrogen, particularly clean hydrogen, plays a critical role in these strategies by providing an alternative energy source that

## Nomenclature

**Indexes:**

| | |
|---|---|
| $I$: | Set of hydrogen types ($i \in$ {green, blue, grey}) |
| $J$: | Set of hydrogen producers ($j \in$ {CHPE, SHP}) |
| $T$: | Set of time periods |

**Input parameters:**

| | |
|---|---|
| $\theta$: | Production marginal cost |
| $d_i^t$: | Potential demand for hydrogen type $i$ under a zero-price scenario assumption |
| $n_0$: | Initial penetration level ($n_0 > 0$) |
| $rt_i$: | Diminishing level of accumulated penetration at period $t$ |
| $\gamma$: | Cost associated with the late delivery time |
| $\rho$: | Cost associated with reducing the late delivery time |
| $lt_j$: | The delivery lead time of hydrogen producer $j$ |
| $q_i$: | Average price of alternative fuels in the market at period $t$ |
| $\beta$: | Coefficient representing the hub's penetration effect on demand |
| $b_0$: | Cross-elasticity effect of the average price of hydrogen type $i$ at period $t$ |
| $s_i^t$ ($\underline{s}_i$): | Sensitivity of demand for hydrogen type $i$ to price (delivery lead time) |
| $\tau$: | Standard delivery lead time expected by customers |
| $k_{it}^{ji}$: | Maximum capacity of hydrogen production facilities $j$ for hydrogen type $i$ achievable at period $t$ |
| $a_i$: | Market penetration sensitivity to green production contribution, $w_t$ |
| $s_{it}^{SHP}$ ($c_{it}^{CHPE}$): | Cost of hydrogen type $i$ produced by the SHP (CHPE) at period $t$ |
| $\delta_i$: | Hub commission imposed on the hydrogen producers at period $t$ |

**Decision variables:**

| | |
|---|---|
| $Q_{it}^{SHP}$ ($Q_{it}^{CHPE}$): | Quantities of hydrogen type $i$ produced by the SHP (CHPE) at period $t$ |
| $r_t$: | Penetration state at time $t$ |
| $w_t$: | Green hydrogen rate within total available hydrogen in the hub at period $t$ |
| $x_{it}^{SHP}$ ($x_{it}^{CHPE}$): | Price offered by the SHP (CHPE) for hydrogen type $i$ at period $t$ |
| $y_t$: | Reduction from CHPE's delivery lead time at period $t$ |
| $\varphi_t$ ($\psi_t$): | Discount factor within the tariff contracts among the hydrogen hub and SHP (CHPE) for hydrogen type $i$ at period $t$ |
| $\omega_t$ ($\bar{\omega}_t$): | Lump-sum payment paid by the SHP (CHPE) for hydrogen type $i$ at period $t$ |

**Abbreviation:**

| | |
|---|---|
| CHPE: | Conventional Hydrogen Production Efficiency |
| SHP: | Sustainable Hydrogen Production |
| RER: | Renewable energy resource |
| GHG: | Greenhouse gases |
| Co: | Coordination Scheme |
| Ct: | Market-Based Pricing Structure |
| Cn: | Cooperative Integration Structure |

can mitigate the environmental impact of the industrial and transportation sectors. This makes hydrogen not just a replacement for fossil fuels but also a cornerstone for achieving long-term sustainability and economic growth in a decarbonized world.

This study explores how hydrogen pricing and supply lead time impact the demand for hydrogen under different collaboration policies, as they directly influence the adoption rate and accessibility of hydrogen fuel, thereby shaping its demand in an increasingly competitive energy market. Pricing, demand, and collaboration determine the feasibility and attractiveness of hydrogen as a sustainable energy alternative, impacting supply dynamics and consumer choice. This study considers a price-sensitive demand rate function for hydrogen. The selling price depends on how environmentally friendly the production methods are, the cost of alternative fuel options, and the fuel substitution dynamics. It also includes environmentally friendly production and the cross-elasticity effect of delivery efficiency in estimating demand, highlighting the challenges and opportunities inherent within the hydrogen landscape.

This study also examines the concept of market penetration. Market penetration measures the proportion of customers adopting a product or service relative to the estimated market share for that product or service [4]. It helps develop strategies to increase the market share of those products or services [5]. Market penetration is also applied in the hydrogen economy to assess hydrogen as a potentially competitive and viable energy alternative [6]. Knowing market penetration enables hydrogen stakeholders to predict changes in what people want and how well-known it is in the market. It also helps identify the growth opportunities and potential barriers in the hydrogen market, guiding resource allocation and investment strategies. It will further help hydrogen producers assess the effectiveness of their penetration strategies compared to others. The literature fails to explore the effects of market penetration on hydrogen demand. While some studies have looked at the impact of market penetration on renewable energy sources (e.g., Refs. [7–9]), few have focused on hydrogen (e.g., Refs. [10,11]). This study fills this gap by thoroughly analyzing market penetration and its effects on maximizing market share, profitability, and long-term sustainability within the hydrogen industry.

Hydrogen producers are either Conventional Hydrogen Production Entities (CHPEs) or Small-scale Hydrogen Producers (SHPs). CHPEs are large-scale and use mostly fossil fuels to produce hydrogen, while SHPs are much smaller-scale and use renewable energy sources instead. This study examines, therefore, the complexities surrounding collaboration strategies among key players in the hydrogen market by analyzing different game structures to explain how CHPE and SHP navigate a dynamic hydrogen market environment.

This study evaluates three different scenarios: (*i*) market-based pricing structure, (*ii*) cooperative integration structure, and (*iii*) coordinated decision-making structure. In a market-based pricing structure, CHPE and SHP compete through a pricing mechanism, driving players' efficiency while meeting consumer demand. Under the cooperative integration structure, CHPE and SHP form strategic alliances to address challenges and ensure sustainability. Coordinated decision-making facilitated by a commission-based structure establishes a centralized hub for CHPE and SHP to align actions and maximize outcomes in the hydrogen economy. These collaborative models, particularly within the framework of hydrogen marketing using a hub, not only facilitate market access but also introduce a commission-based incentive structure for the hub, aligning interests across the hydrogen value chain.

The importance of this study is that it analyzes factors influencing demand for clean hydrogen and evaluates collaboration models that will help us better understand the dynamics of a hydrogen economy and its complexities. It will provide practical insights on pricing and strategic partnerships, demand shifts and consumer preferences, and the transition into a more sustainable, competitive, and growing hydrogen economy. It will also investigate the role of the hydrogen hubs in increasing the market share for hydrogen and the profitability of those who produce and trade it. However, it is crucial to recognize the uncertainties and potential barriers, such as fluctuating production costs and technological advancements, that may impact the transition to clean hydrogen. The varying bargaining powers of each member within the collaborative framework also merit investigation to comprehend their roles in fostering an environment conducive to sustainable growth. Lastly, this study answers how policymakers can influence and facilitate the transition to cleaner hydrogen production practices that align with their environmental objectives.

## 2. Literature review

The literature on hydrogen energy is rich and covers topics from production methods to market dynamics. The review in this section focuses only on hydrogen pricing, market penetration, and collaboration strategies, areas relevant to this study.

### 2.1. Pricing and factors influencing demand for hydrogen

A green hydrogen economy is the answer to achieving zero carbon emissions in multiple sectors [12]. Wolfram et al. [13] showed that clean hydrogen deployment can reduce total energy decarbonization costs by 15%–22 %.

The cost of hydrogen production varies with the method and energy source used. The cost of producing hydrogen from conventional sources, like natural gas or coal, is projected to remain relatively the same between now and 2050 [12]; green hydrogen production costs are generally higher due to the electricity cost needed for electrolysis, though those costs are expected to reduce over time (e.g., Refs. [13,14]). For instance, in Canada, where 82 % of the electricity supply comes from renewable or non-GHG-emitting sources, the production of green hydrogen has enormous growth potential [15]. In Australia, the cost of producing a kilogram (kg) of green hydrogen is about \$3.18–3.80 in 2024 and will be \$2 by the end of the decade [16]. This happens as the technology matures, with the cost per kg reducing following some form of a learning curve. There is ample empirical evidence that the costs of renewable energy technologies (e.g., solar, wind, hydropower) and, subsequently, that of generating electricity influence hydrogen pricing [17].

Despite the expected reduction in the cost to produce a kg of hydrogen, its pricing remains complex due to other factors such as production methods, governmental subsidies, and taxes [18]. There are three hydrogen production methods. Green hydrogen is produced using renewable energy sources. Blue and grey hydrogen production uses natural gas, with and without carbon capture technology, respectively [19]. Governmental policies such as taxes on non-green hydrogen production and green hydrogen sales subsidies significantly influence pricing [18]. Studies like the one by Gulli et al. [1] projected the competitiveness of green hydrogen prices with grey hydrogen in the coming years. How pricing dynamics interact with delivery lead time to influence demand still needs to be understood. This study addresses the above gap by incorporating production methods on selling price, the cost of alternative fuels, and substitution dynamics.

Demand for hydrogen is highly price sensitive [1]. Consumers tend to move away from hydrogen to alternative fuels despite the energy savings that could be greater than the price differential [20]. Some research has shown that future demand for non-green hydrogen, such as grey hydrogen, is mainly influenced by cost savings compared to conventional fuels like gasoline [21]. Although the relationship between pricing dynamics and demand for other products has been investigated by several studies (e.g., Ref. [22]), there remains a gap in the literature on identifying changes in customers' preferences. This study uses an extension of the existing function and addresses this gap by presenting a price-sensitive demand function that considers the price of alternative fuels. The demand function aims to model customer behavior regarding hydrogen demand.

As the cost of producing one kg of hydrogen reduces, a large percentage of the unit cost will be for distributing and delivering it to customers [23]. This delivery impacts the demand for hydrogen [20]. "Lead time" refers to the period between placing an order and receiving the hydrogen. When delivery lead times are shorter, customers experience greater convenience and flexibility, leading to increased adoption of hydrogen as an energy carrier. Conversely, longer lead times can deter potential users, especially in time-sensitive applications such as industrial processes or transportation. Delivery time is crucial for these sectors because the timely availability of hydrogen directly impacts operational efficiency, where delays can result in increased costs and decreased productivity. There are a few studies on hydrogen delivery efficiency in the literature. For example, O'Rourke et al. [24] evaluated the most popular hydrogen production methods in conjunction with delivery options.

This study contributes to this field by examining the cross-elasticity effect of delivery lead time, specifically, how deviations from the standard expected lead time influence customer preferences and overall hydrogen demand. As this study considers these dynamics, it aims to enhance the understanding of supply chain efficiency and customer satisfaction in the hydrogen economy.

### 2.2. Collaboration strategies in the hydrogen market

The transition to a net-zero economy necessitates collaborative efforts across various stakeholders. Many countries have released national hydrogen strategies, emphasizing collaboration among governments, industry players, and research institutions. These strategies outline key focus areas, including exports, imports, and technological leadership, which inherently involve collaboration and coordination. Reports from organizations like KPMG [25], Fortune Business Insights [26], and Maximize Market Research [27] discuss hydrogen market dynamics, investment trends, and technological advancements. While not exclusively focused on collaboration, these reports often mention the need for coordinated efforts to address challenges and seize opportunities in the hydrogen sector. For instance, the First Movers Coalition exemplifies how companies can join forces to accelerate complex and capital-intensive hydrogen projects, bridging supply and demand gaps [28]. Incorporating considerations for hydrogen hubs into these discussions is also necessary. Collaboration can unlock investment and drive hydrogen market growth through fostering partnerships, sharing risks, and aligning interests.

Although studies examining the complexities surrounding collaboration strategies among key players in the hydrogen market may be limited, the broader discourse on hydrogen market development consistently underscores the need for game-theoretic approaches to analyze complex interactions among independent players in the hydrogen market. While studies such as the one conducted by Huang and Li [29] explored the application of game theory in other energy systems, no study optimizes collaboration strategies to gain the largest share of the hydrogen market. This study bridges this gap by analyzing the dynamics between two distinct hydrogen entities, i.e., CHPE and SHP, under different market conditions, i.e., Cournot Competition, Collusion, and Tariff Contracts.

Cournot competition is frequently used to model how firms decide on production quantities while considering competitors' reactions and maximizing individual profits. In the context of the hydrogen market, where producers are interdependent, Cournot-based models highlight the tension between competition and cooperation. This is particularly relevant when considering hydrogen hubs, where producers interact within a shared infrastructure, influencing each other's profits [30]. In a collusion game, which introduces cooperative settings to the discussion, hydrogen producers may form coalitions to maximize joint profits by aligning pricing and production decisions. The literature on collusion and cartel behavior in energy markets offers insights into how such arrangements can lead to higher profitability than competitive scenarios (e.g., Refs. [31,32]). Hydrogen market collusion analyses, particularly through tariff contracts, extend this theory by incorporating specific policy mechanisms to facilitate cooperation while balancing regulatory oversight. Finally, tariff contracts are designed to ensure that collaboration among hydrogen producers is profitable and sustainable over time [33]. The acceptance conditions derived in this study ensure that all parties find the contract terms beneficial, which aligns with the broader game-theoretic literature on incentive compatibility and contract enforcement in collaborative environments. This study builds upon these theoretical foundations by employing game theory frameworks to optimize decision-making among interacting and competing bodies. By analyzing these interactions within a Cournot competition and a Collusion framework, this study contributes valuable insights into analyzing hydrogen markets and the role of collaboration in achieving a competitive advantage.

## 3. Material and methods

This study considers a system consisting of two hydrogen producers, CHPE and SHP, supplying a hydrogen hub that sells hydrogen to customers for different uses (see Fig. 1). The hub coordinates sales to domestic customers. It ensures efficient distribution and delivery and manages producer collaboration to optimize logistics. The hub handles liquefaction or conversion to ammonia, facilitating cost-effective exports. The CHPE uses coal and natural gas to generate electricity to produce (Grey) hydrogen and emits $CO_2$. The SHP uses renewable energy and natural gas to produce electricity and emits some $CO_2$ in the process. The hydrogen produced from this process is labeled as 'Blue' hydrogen. An SHP that generates electricity from a renewable resource to produce (Green) hydrogen emits no $CO_2$. The water that CHPE and SHP consume and release is not contaminated and is considered safe, although some systems should be in place to ensure water quality. Sustainability is a collective effort that involves government, producers, distributors, and consumers. Consumers' participation and commitment to green choices are necessary for the success of a sustainable initiative. However, hydrogen delivery lead times and cost-effectiveness could alter consumer preferences that either help the demand for hydrogen to grow or slow it.

In Fig. 1, the CHPE and the SHP can compete following a Cournot game or cooperate following a collusion game. A Cournot competition involves companies competing by simultaneously determining output levels, while collusion represents a form of cooperative game theory where companies coordinate to restrict output and attain monopoly prices. Solving these games will help us understand how hydrogen production decisions affect market behavior and dynamics. This study examines what drives demand for hydrogen by considering environmental sustainability, economic viability, and strategic decision-making among industry stakeholders.

### 3.1. Assumptions

**Assumption 1.** The hydrogen production quantities at time t cannot exceed the production capacities of CHPE and SHP, i.e., $Q_{it}^{CHPE} < k_{it}^{CHPE}$ and $Q_{it}^{SHP} < k_{it}^{SHP}$.

**Assumption 2.** CHPE generally generates hydrogen at a lower cost than SHP ($x_{it}^{CHPE} < x_{it}^{SHP}$). CHPE often uses mature methods like steam methane reforming to produce hydrogen, allowing them to benefit from economies of scale [30], i.e., the cost per kg of hydrogen reduces as the quantity of produced hydrogen increases. SHP often uses more expensive methods (e.g., electrolysis) to produce hydrogen [31]. SHP has the advantage of supplying smaller but more frequent quantities due to their proximity to the point of use, which reduces the hydrogen delivery time ($lt^{SHP} < lt^{CHPE}$).

**Assumption 3.** The more services a hydrogen hub offers, the more customers it will reach. This study investigates how the method used to produce hydrogen and hydrogen prices affects the speed of penetrating an energy market. In simpler terms, as the hub, consists of $I$ hydrogen types (i.e., green, blue, and grey) and $J$ hydrogen producers (i.e., CHPE and SHP), increases its share of environmentally friendly hydrogen production, and offers lower prices, customers tend to be more satisfied using hydrogen from this platform. In this context, renewable energy resources (RERs) that have more share in producing hydrogen tend to have better chances to penetrate the energy market, which implies higher customer satisfaction and market competitiveness for the hydrogen hub. Also, high prices hurt consumers and, subsequently, profit margins, lowering the economic sustainability of a hydrogen hub. Building consumer awareness of purchase options and environmentally friendly production methods also takes time. The study considers these dynamic effects over time, defining penetration as a state variable

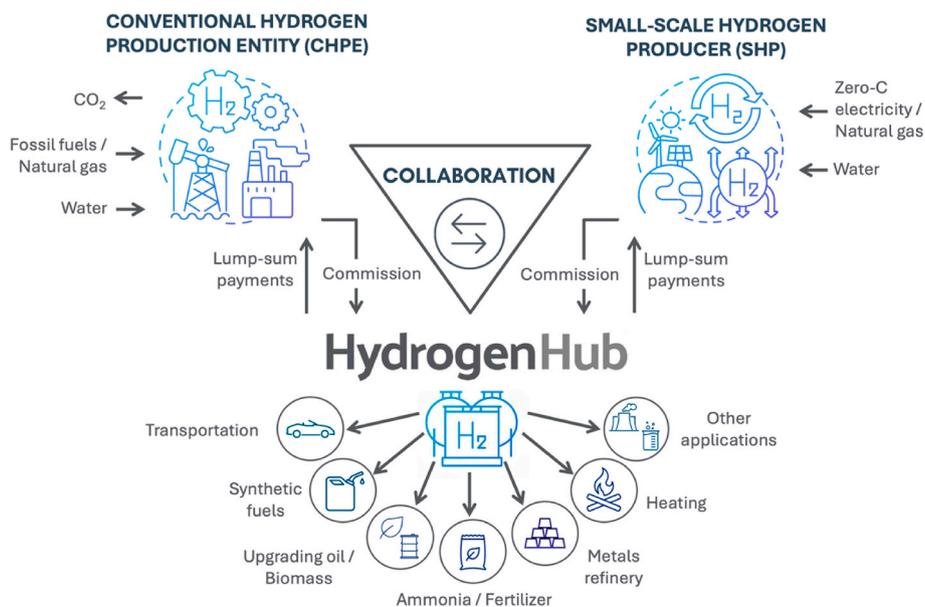

**Fig. 1.** Structure of the analyzed hydrogen market.

reflecting its status at the next time, formulated as:

$$r_{t+1} = (1 - rl_t) \cdot \left[ r_t + \alpha_t \cdot w_t - \sum_{i \in I} b_{it} \cdot \ln(\bar{x}_{it}) \right] \quad (1)$$

Here, $\bar{x}_{it} = \frac{x_{it}^{CHPE} + x_{it}^{SHP}}{2}$ represents the average price offered by the CHPE and SHP for a kg of hydrogen type $i$ (i.e., $i$ = green, blue, and grey) at time $t$. The diminishing level, $rl_t$, assumes that the penetration level implies a slower demand for hydrogen as time passes and the market becomes saturated and competitive.

**Assumption 4.** This study accounts for pricing, the sensitivity of customers to prices of alternative fuels, and the speed of delivery to investigate changes in demand for hydrogen, penetrating the energy market, and the market behavior. This approach extends a general function proposed by Asghari et al. [22], who initially explored the relationship between pricing dynamics and demand while considering other products available in the market. The possibility of substituting one type of hydrogen for another is denoted by $\vartheta \epsilon (0,1)$. Considering the offered price for hydrogen type $i$ ($x_{it}^j$) and reduction in delivery lead time ($y_t^j$), the demand for different types of hydrogen produced by both entities $j \in \{SHP, CHPE\}$ at period $t$ can be formulated as follows:

$$Q_{it}^j = d_i^0 \cdot \left[ \beta \cdot r_t + e^{-s_1^i \cdot \left( \frac{x_{it}^{j\,\vartheta+1}}{\vartheta_t^{\vartheta^\vartheta}} \right) - s_2^i \left( \frac{lt^j - y_t^j}{\tau} \right)} \right] \quad (2)$$

where $d_i^0$ denotes the maximum amount of hydrogen type $i$ that can be consumed across all sectors in an extreme scenario when prices are set to zero. This value represents the potential demand without accounting for the influence of other parameters. In Eq. (2), the parameters $s_1^i$ and $s_2^i$ represent the price and delivery time sensitivities, respectively, for hydrogen type $i$; this is because consumers have more loyalty to the green hydrogen coming from renewable or non-GHG-emitting sources and are affected less by price and delivery time. After calculating $Q_{it}^j$, $w_t$ is determined by dividing the combined green hydrogen quantities produced by CHPE and SHP ($Q_{1t}^{SHP} + Q_{1t}^{CHPE}$) by the total hydrogen production ($\sum_{j \in J} \sum_{i \in I} Q_{it}^j$).

### 3.2. Formulation of the profit dynamics

CHPE's lower hydrogen prices, compared to SHP, significantly affect price-sensitive customers. Since SHPs mostly use renewable energy and new technology, their unit hydrogen production costs increase with more hydrogen produced. Unlike conventional technologies, renewable energy technologies typically do not benefit much from economies of scale even when they mature. Thus, as these technologies improve, the costs of operating them reduce. Of course, the steeper the learning curve for these technologies, the sooner they become more appealing and competitive. Although this study does not consider SHP learning curves, it opens the door for a future research path. Thus, it is reasonable to assume that the cost of producing a kg of hydrogen increases following a diminishing returns function with parameter $\theta$ ($\theta$ has dimensions of cost per square ton of hydrogen, i.e., $/ton^2$), i.e., $\theta Q_{it}^2$, with $Q_{it}^2$ representing the growing marginal cost. An SHP supplying hydrogen type $i$ (blue or green) at $t$ incurs a cost of $\theta Q_{it}^2$ (in Eq. (3) below) that increases with the quantity (in tons) of hydrogen produced and sold. This economic principle is referred to as the Law of Diminishing Returns proposed by Shephard and Färe [32]. This concept is widely used in production theory and plays a central role at the micro- and macroeconomics levels. It helps to explain why, beyond a certain point, it becomes less efficient to continue increasing production and more cost-effective to explore other options, such as improving technology or optimizing the production process.

Typically, there is a predetermined time window for delivery. The challenge lies in the variability of this window. Shipments arriving ahead of schedule incur additional costs. A penalty is incurred when it arrives after the designated time window (see Ref. [33]). CHPE invests $\rho$ dollars for a delivery lead-time of $y_t$ days. The investment cost is affected by $\rho$, $y_t$, and $Q_{it}$, i.e., $\rho \cdot y_t \cdot \log_{\xi}(Q_{it} + 1)$ in Eq. (4) below, where the rate of increasing this cost slows with larger quantities, suggesting a form of economics of scale. Very short or long lead times result in losing customers and disrupting the energy market. Repetitive incidents of lengthy delivery times cause customer dissatisfaction to grow faster, more exponentially than linearly. This constitutes the cost $\gamma \cdot \left[ (lt - y_t) - \tau \right]^2$, where $\gamma$ is a delivery lead time parameter. This parameter escalates rapidly beyond the expected delivery time $\tau$ days to emphasize the importance of timely deliveries. So, the investment amounts to reducing the delivery lead time in Eq. (4) is $\gamma \cdot \left[ (lt - y_t) - \tau \right]^2 + \rho \cdot y_t \cdot \log_{\xi}(Q_{it} + 1)$, note that adding 1 ensures the function is adjusted for $Q_{it}^{CHPE} = 0$. The CHPE markets its hydrogen through a hub, and in this arrangement, the hub earns a commission for each ton of hydrogen sold through the platform. At each period, the hub imposes a commission denoted as $\delta_t$ on the CHPE and SHP. Consequently, the profit functions at period $t$ can be computed as follows:

$$\pi_{it}^{SHP}(x_{it}) = (1 - \delta_t) \cdot \left[ x_{it}^{SHP} - c_{it}^{SHP} \right] \cdot Q_{it}^{SHP} - \theta \cdot Q_{it}^{SHP\,2} \quad (3)$$

$$\pi_{it}^{CHPE}(x_{it}, y_t) = (1 - \delta_t) \cdot \left[ x_{it}^{CHPE} - c_{it}^{CHPE} \right] \cdot Q_{it}^{CHPE} - \gamma \cdot (lt - y_t - \tau)^2 \\ - \rho \cdot y_t \cdot \log_{\xi}(Q_{it}^{CHPE} + 1) \quad (4)$$

where $c_{it}^{SHP}$ and $c_{it}^{CHPE}$ are the cost of hydrogen produced by the CHPE and SHP, respectively. The term $\left[ x_{it}^j - c_{it}^j \right] \cdot Q_{it}^j$, where $j$ = SHP or CHPE, is the net revenue for hydrogen producer $j$. The first term of Eq. (4) encapsulates the CHPE's revenue, while the second term represents the investment allocated to reducing the delivery time, which decreases at an increasing rate with $Q_{it}^{CHPE}$. The profit function for hydrogen hub becomes:

$$\pi_{it}^{HUB}(Q_{it}) = \delta_t \cdot \sum_{j \in J} \left( P_{it}\left( Q_{it}^j \right) - c_{it}^j \right) \cdot Q_{it}^j \quad (5)$$

Herein, $P(Q)$ is the inverse demand function ($x^j = P_{(t)}(Q^j)$) as indicated in Appendix A. The inverse of the demand function is calculated to derive the price based on the total hydrogen produced. This approach is grounded in established economic principles, where increased demand, given a limited supply, results in higher prices.

Three collaboration models for the system described in Fig. 1 are developed for market-based pricing, cooperative integration, and coordinated decision-making structures.

#### 3.2.1. Market-based pricing in Cournot game

Market-based pricing is a strategy that determines prices based on market demand and supply dynamics. From the perspective of hydrogen production, this strategy can be modeled using a Cournot game. In a Cournot game, each hydrogen producer independently and simultaneously selects its output level, considering the outputs of other producers as constants. Each producer makes its decision based on the assumption that other hydrogen producers' decisions are not affected by its own. The output of all hydrogen producers dictates prices. The profit of the CHPE depends on its output and that of the SHE, and vice versa. In the Cournot game, denoted by the superscript Ct, the SHE determines its market price and the discount rate. The CHPE simultaneously sets its hydrogen prices and makes investment decisions. To enhance resilience and environmental sustainability in hydrogen production and identify optimal decision factors, members initially optimize their variables independently, aiming to derive closed-form solutions.

**Proposition 1.** The profit functions for both hydrogen producers, CHPE and SHP, exhibit concavity with respect to their respective quantities produced ($Q_{it}$) when following Cournot behavior.

Proof **of Proposition 1.** Since $\theta > 0$, $\rho > 0$, $Q_{it}^{SHP} \geq 0$, and $Q_{it}^{CHPE} \geq 0$, the second-order derivatives of the SHP's and CHPE's profit under Cournot game structure are $\frac{\partial^2 \pi_{it}^{SHP}(Q_{it}^{SHP} + Q_{it}^{CHPE})}{\partial Q_{it}^{SHP^2}} = -2\theta < 0$ and

$P_{it}\left(Q_{it}^{SHP} + Q_{it}^{CHPE}\right)$. Thus, by replacing the optimal values of hydrogen production, $Q_{it}^{j_{Co}}$ (where $j$ belongs to either CHPE or SHP), into the profit function of the hub, the non-cooperative decision-making model under the Cournot behavior of both CHPE and SHP, denoted as $NCM^{Ct}$, is formulated as follows:

$$NCM^{Ct} : \max \pi_{it}^{HUB}\left(Q_{it}^{SHP_{Co}}, Q_{it}^{CHPE_{Co}}\right) = \delta_t \cdot \left[\left(P_{it}\left(Q_{it}^{SHP} + Q_{it}^{CHPE_{Co}}\right) - c_{it}^{SHP}\right) \cdot Q_{it}^{SHP_{Co}} + \left(P_{it}\left(Q_{it}^{SHP} + Q_{it}^{CHPE_{Co}}\right) - c_{it}^{CHPE}\right) \cdot Q_{it}^{CHPE_{Co}}\right] \tag{9}$$

$\frac{\partial^2 \pi_{it}^{CHPE}(Q_{it}^{SHP} + Q_{it}^{CHPE})}{\partial Q_{it}^{CHPE^2}} = -\rho \cdot y_t \cdot \frac{1}{\left(Q_{it}^{CHPE} + 1\right)^2} < 0$, respectively. Thus, both $\pi_{it}^{SHP}$ and $\pi_{it}^{CHPE}$ are concave with respect to $Q_{it}$, where $Q_{it} = Q_{it}^{SHP} + Q_{it}^{CHPE}$. By setting $\frac{\partial \pi_{it}^{SHP}(Q_{it})}{\partial Q_{it}^{SHP}} = 0$ and $\frac{\partial \pi_{it}^{CHPE}(Q_{it})}{\partial Q_{it}^{CHPE}} = 0$, the optimal values of hydrogen production by both producers are (see Appendix B):

$$Q_{it}^{SHP_{Co}} = \frac{s_1^t \cdot (\vartheta + 1) \cdot \left[Q_{it}^{CHPE} - d_{it}^0 \cdot \beta \cdot r_t\right]}{q_t^\vartheta}$$

$$+ e^{\left(\frac{s_1^t}{q_t^\vartheta}\left(\frac{2\theta}{1-\delta_t} \cdot q_t^{\vartheta\vartheta}\right)^{\vartheta+1} - \frac{s_1^t}{s_1^t} \cdot q_t^\vartheta \cdot \left(\frac{lt}{\tau}\right) - \ln(d_{it}^0) + \epsilon_1\right)} \tag{6}$$

$$Q_{it}^{CHPE_{Co}} = \frac{s_1^t \cdot (\vartheta + 1) \cdot \left[Q_{it}^{SHP} - d_{it}^0 \cdot \beta \cdot r_t\right]}{q_t^\vartheta}$$

$$+ e^{\left(\frac{\left[c_{it}^{CHPE} - \frac{\rho \cdot y_t}{1-\delta_t}\right]^{\vartheta+1} - \frac{s_1^t}{s_1^t} \cdot q_t^\vartheta \cdot \left(\frac{lt - y_t}{\tau}\right) - \ln(d_{it}^0) + \epsilon_2}{\frac{q_t^\vartheta}{s_1^t}\left(Q_{it}^{SHP} - d_{it}^0 \cdot \beta \cdot r_t\right)}\right)} \tag{7}$$

*s.t.*

$$Q_{it}^{SHP_{Co}} \in argmax \ \pi_{it}^{SHP}(Q_{it})$$

$$Q_{it}^{CHPE_{Co}} \in argmax \ \pi_{it}^{CHPE}(Q_{it})$$

### 3.2.2. Cooperative integration in collusion game

Cooperative integration is a strategy where multiple entities work together to achieve a common goal. From the perspective of a hydrogen production, this can be modeled using a Collusion game. Under Collusion behavior, indicated by (Cn), both hydrogen producers aim to jointly optimize the selling prices of various hydrogen types to maximize the total profit $\pi_{it}^{SHP} + \pi_{it}^{CHPE}$. This will result in a higher collusive quantity than in the non-collusive equilibrium. The combined profit function at period $t$ under the Collusion game, which is denoted as $CPF^{Cn}$, can be defined as follows:

$$CPF^{Cn} : \max \pi_{it}^T\left(Q_{it}^{SHP}, Q_{it}^{CHPE}\right) = (1 - \delta_t) \cdot \left[\left(x_{it}^{SHP} - c_{it}^{SHP}\right) \cdot Q_{it}^{SHP} + \left(x_{it}^{CHPE} - c_{it}^{CHPE}\right) \cdot Q_{it}^{CHPE}\right] - \gamma \cdot (lt - y_t - \tau)^2 - \theta \cdot Q_{it}^{SHP^2} - \rho \cdot y_t \cdot \log\left(Q_{it}^{CHPE} + 1\right) \tag{10}$$

These two equations determine the Cournot equilibrium quantities $Q_{it}^{SHP_{Co}}$ and $Q_{it}^{CHPE_{Co}}$. To find the optimal reduction in delivery lead time ($y_t^{Ct}$) for optimal results in the Cournot game, the first derivative of the profit function for CHPE with respect to $y_t$ is set equal to zero (i.e., $\frac{\partial \pi_{it}^{CHPE}(Q_{it}^{SHP} + Q_{it}^{CHPE})}{\partial y_t} = 0$). This is outlined in detail in Appendix C. After isolating the variable on one side of the equation, its value is calculated as follows:

$$y_t^{Ct} = \frac{lt - \tau - \frac{1}{2\gamma}\left(D(1) \cdot \left[-\frac{s_1^t}{s_1^t} \cdot q_t^\vartheta \cdot \left(\frac{lt}{\tau} - 1\right) - D(2)\right]^{\frac{-\beta}{\beta+1}} - \rho \cdot \log\left(Q_{it}^{CHPE} + 1\right)\right)}{1 + \frac{1}{2\gamma}\left(D(1) \cdot \left(\frac{s_1^t}{s_1^t} \cdot q_t^\vartheta\right) \cdot \left(\frac{-\beta}{\beta+1}\right) \cdot \left[-\frac{s_1^t}{s_1^t} \cdot q_t^\vartheta \cdot \left(\frac{lt}{\tau} - 1\right) - D(2)\right]^{\frac{-\beta}{\beta+1}}\right)} \tag{8}$$

The terms $D(1)$ and $D(2)$ referenced in Eq. (8) are explained and defined in detail in Appendix C.

The Cournot equilibrium for hub penetration can be calculated using variables $w_t^{Ct} = \frac{Q_{it}^{SHP_{Co}} + Q_{it}^{CHPE_{Co}}}{\sum_{j \in J} \sum_{i \in I} Q_{it}^{Co}}$ and $y_t^{Ct}$. The equilibrium price is then

**Proposition 2.** In the Collusion game, the $CPF^{Cn}$ function demonstrates concavity concerning the quantities produced, namely $Q_{it}^{SHP}$ and $Q_{it}^{CHPE}$.

Proof **of Proposition 2.** The second-order derivatives of the combined profit under Cournot game structure with respect to $Q_{it}^{SHP}$ and $Q_{it}^{CHPE}$ are $\frac{\partial^2 \pi_{it}^T(Q_{it}^{SHP}, Q_{it}^{CHPE})}{\partial Q_{it}^{SHP^2}} = -2\theta < 0$ and $\frac{\partial^2 \pi_{it}^T(Q_{it}^{SHP}, Q_{it}^{CHPE})}{\partial Q_{it}^{CHPE^2}} = -\rho \cdot y_t \cdot \frac{1}{\left(Q_{it}^{CHPE} + 1\right)^2} < 0$, respectively. Thus, $CPF^{Cn}$ is concave with respect to both $Q_{it}^{SHP}$ and $Q_{it}^{CHPE}$.

By setting $\frac{\partial \pi_{it}^T(Q_{it}^{SHP}, Q_{it}^{CHPE})}{\partial Q_{it}^{SHP}} = (1 - \delta_t) \cdot \left(x_{it}^{SHP} - c_{it}^{SHP}\right) - 2 \cdot \theta \cdot Q_{it}^{SHP} = 0$ and $\frac{\partial \pi_{it}^T(Q_{it}^{SHP}, Q_{it}^{CHPE})}{\partial Q_{it}^{CHPE}} = (1 - \delta_t) \cdot \left(x_{it}^{CHPE} - c_{it}^{CHPE}\right) - \frac{\rho \cdot y_t}{Q_{it}^{CHPE} + 1} = 0$, the optimal values of hydrogen production by both entities, as indicated in Appendix D, can be calculated by Eqs. (11) and (12).

$$Q_{it}^{SHP_{Cn}} = d_{t0}'.\beta.r_t + e^{\left(\frac{\frac{s_2'}{q_2^\gamma}\left(\frac{2.\theta}{1-\delta_t}+c_{it}^{SHP}\right)^{\beta+1}-\frac{s_2'}{q_2^\beta}.q_s^\beta.\left(\frac{lt}{\tau}\right)-\ln(d_{t0}')+\varepsilon_1}{\frac{s_2'}{q_2^\beta}}\right)} \tag{11}$$

$$Q_{it}^{CHPE_{Cn}} = d_{t0}'.\beta.r_t + e^{\left(\frac{\left(c_{it}^{CHPE}-\frac{\rho.y_t}{1-\delta_t}\right)^{\beta+1}+\frac{s_1'}{q_1^\beta}.q_s^\beta.\left(\frac{lt-y_t}{\tau}\right)-\ln(d_{t0}')+\varepsilon_2}{\frac{q_s^\beta}{s_1'}}\right)} \tag{12}$$

These two equations establish the optimal quantities, $Q_{it}^{SHP_{Cn}}$ and $Q_{it}^{CHPE_{Cn}}$, representing the equilibrium in Collusion. As $\pi_{it}^{SHP}$ in $\pi_{it}^T = \pi_{it}^{SHP} + \pi_{it}^{CHPE}$ does not contain the variable $y_t$, the optimal reduction in delivery lead time in Collusion game can be calculated by setting $\frac{\partial \pi_{it}^{CHPE}(Q_{it}^{CHPE})}{\partial y_t} = 0$. The optimal value of $y_t^{Cn}$ is calculated similar with Cournot as:

$$y_t^{Cn} = \frac{lt - \tau - \frac{1}{2\gamma}\left(D(1).\left[-\frac{s_1'}{s_1'}.q_s^\beta.\left(\frac{lt}{\tau}-1\right)-F(1)\right]^{\frac{-\beta}{\beta+1}}-\rho.\log(Q_{it}^{CHPE}+1)\right)}{1 + \frac{1}{2\gamma}\left(D(1).\left(\frac{s_1'}{s_1'}.q_s^\beta\right).\left(\frac{-\beta}{\beta+1}\right).\left[-\frac{s_1'}{s_1'}.q_s^\beta.\left(\frac{lt}{\tau}-1\right)-F(1)\right]^{\frac{-\beta}{\beta+1}}\right)} \tag{13}$$

Here $F(1) = \frac{q_s^\beta}{s_1'}.\ln\left(\frac{Q_{it}^{CHPE}-d_{t0}'.\beta.r_t}{d_{t0}'}\right)$. Subsequently, $w_t^{Cn}$ and $r_{t+1}^{Cn}$ can be calculated with these variables. By substituting the optimal values of hydrogen production, $Q_{it}^{j_{Cn}}$ (where $j$ can be either CHPE or SHP), into the profit function of the hub, the cooperative decision-making model under the Collusion behavior of both CHPE and SHP, denoted as $CM^{Cn}$, is formulated as:

$$CM^{Cn}: \max \pi_{it}^{HUB}(Q_{it}^{SHP_{Cn}}, Q_{it}^{CHPE_{Cn}}) = \delta_t.\left[\left(P_{it}(Q_{it}^{SHP}) - c_{it}^{SHP}\right).Q_{it}^{SHP_{Cn}} + \left(P_{it}(Q_{it}^{CHPE}) - c_{it}^{CHPE}\right).Q_{it}^{CHPE_{Cn}}\right] \tag{14}$$

$s.t.$

$$\left(Q_{it}^{SHP_{Cn}}, Q_{it}^{CHPE_{Cn}}\right) \in argmax\ \pi_{it}^T(Q_{it}^{SHP}, Q_{it}^{CHPE})$$

### 3.2.3. Coordinated decision-making structure

In the proposed hydrogen production scheme, the efficiency of the hub platform is compromised by the competitive dynamics among hydrogen producers and the independent decision-making of each participant. This misalignment of interests necessitates the introduction of coordination tariff contracts to establish connections between SHP and the hydrogen hub and between CHPE and the hydrogen hub. Coordinating the CHPE-hydrogen hub link requires a cost-sharing contract in place. Under this contract, the hydrogen hub contributes to the costs associated with reducing delivery lead time, utilizing a coefficient denoted as $\varphi_{it}'$. This approach is designed to secure the acceptance of centralized decisions by established industry leaders in hydrogen production that are traditionally reliant on conventional methods. The objective is to increase profits for the CHPE and the hydrogen hub compared to the non-integrated option. Coordinating the link between the SHP and the hydrogen hub requires compensating the producer to boost its production capacity, particularly green hydrogen. The associated coefficient, denoted as $\varphi_{it}$ for the SHP fee, is strategically determined to incentivize the active participation of SHP in the coordination system. As part of this coordination scheme, denoted as (Co), managers of the hub charge a lump-sum payment of $\omega_{it}$ and $\omega_{it}'$ from the SHP and CHPE, respectively.

Within this coordinated framework, the profit functions of the involved entities are modified accordingly, aiming to achieve a more efficient and collaborative hydrogen production system. The profit functions are as follows:

$$\pi_{it}^{SHP_{Co}}(x_{it}^{Cn}) = (1 - \delta_t).\left[x_{it}^{SHP} - c_{it}^{SHP}\right].Q_{it}^{SHP_{Co}} - (1 - \varphi_{it}).\theta.Q_{it}^{SHP_{Co}2} - \omega_{it} \tag{15}$$

$$\pi_{it}^{CHPE_{Co}}(x_{it}^{Cn}, y_t^{Cn}) = (1 - \delta_t).\left[x_{it}^{CHPE} - c_{it}^{CHPE}\right].Q_{it}^{CHPE_{Co}} - \gamma.(lt - y_t^{Cn} - \tau)^2 - (1 - \varphi_{it}').\rho.y_t^{Cn}.\log(Q_{it}^{CHPE_{Co}} + 1) - \omega_{it}' \tag{16}$$

$$\pi_{it}^{HUB_{Co}}(Q_{it}) = \delta_t.\left[(P_{it}(Q_{it}^{SHP}) - c_{it}^{SHP}).Q_{it}^{SHP_{Co}} + (P_{it}(Q_{it}^{CHPE_{Co}}) - c_{it}^{CHPE}).Q_{it}^{CHPE_{Co}}\right] - \varphi_{it}.\theta.Q_{it}^{SHP_{Co}2} - \varphi_{it}'.y_t^{Cn}.\log(Q_{it}^{CHPE_{Co}} + 1) + \omega_{it} + \omega_{it}' \tag{17}$$

**Proposition 3.** *The tariff contracts, applied to the competitive connections between SHP and the hydrogen hub and between CHPE and the hydrogen hub, successfully align motivations in the hydrogen production scheme only under the following conditions:*

$$\varphi_{it} = \frac{c_{it}^{SHP} - (1 - \delta_t).x_{it}^{SHP} + \delta_t.c_{it}^{SHP}}{2.\theta.Q_{it}^{SHP_{Co}}} \tag{18}$$

$$\varphi_{it}' = 1 - \frac{(1 - \delta_t).(x_{it}^{CHPE_{Cn}} - c_{it}^{CHPE}).\log \xi.Q_{it}^{CHPE_{Co}} + 1}{\rho.y_t^{Cn}} \tag{19}$$

The proof of Proposition 3 is provided in Appendix E.

The acceptance of the proposed coordination tariff contracts in the competitive connections by hydrogen producers and hub managers hold only when each participant secures profits at least equivalent to those in the Cournot game (market-based pricing approach) and $\pi_{it}^{j_{Co}} \geq \pi_{it}^{j_{Cn}}$ for $j \in \{SHP, CHPE, HUB\}$. Simplifying these conditions allows determining the lower and upper bounds for the lump-sum payments, see. However, the exact values of these parameters are determined through a profit-sharing mechanism, where the surplus profit from the coordination contracts is divided based on the bargaining power of each member.

Various methodologies have been employed to estimate surplus profit sharing, as observed in Chaharsooghi and Heydari [34] and Modak et al. [35]. In this study, surplus profit ($\Delta\pi = \pi^{Cn} - \pi^{Ct}$) is determined by subtracting the profit from a non-cooperative decision-making structure under the Cournot game from that unified profit under the Collusion game.

Considering the bargaining powers of the members as $z^j$ for $j \in \{SHP, CHPE, HUB\}$, the portions of the surplus profits obtained by the CHPE and SHP are $\Delta\pi.\left(\frac{z^{SHP}}{z^{SHP} + z^{CHPE} + z^{HUB}}\right)$ and $\Delta\pi.\left(\frac{z^{CHPE}}{z^{SHP} + z^{CHPE} + z^{HUB}}\right)$, respectively. By employing these specified values, the profit functions within the suggested coordinated models can be expressed as follows:

$$\pi_{it}^{SHP_{Co}}(x_{it}^{Cn}) = \pi_{it}^{SHP_{Co}}\left(P_{it}(Q_{1t}^{SHP} + Q_{1t}^{CHPE}), Q_{it}^{SHP_{Co}}\right) + \Delta\pi.\left(\frac{z^{SHP}}{z^{SHP} + z^{CHPE} + z^{HUB}}\right) \tag{20}$$

$$\pi_{it}^{CHPE_{Co}}(x_{it}^{Cn}, y_t^{Cn}) = \pi_{it}^{CHPE_{Co}}\left(P_{it}(Q_{11t}^{SHP} + Q_{1t}^{CHPE}), Q_{it}^{CHPE_{Co}}, y_t^{Ct}\right) + \Delta\pi.\left(\frac{z^{CHPE}}{z^{SHP} + z^{CHPE} + z^{HUB}}\right) \tag{21}$$

Using Eqs. (20) and (21), the specific values of the lump-sum payments $\omega_{it}$ and $\omega_{it}'$, ensuring the effectiveness of coordination contracts, are determined in the following manner:

$$\omega_{it} = (1 - \delta_t).\left[\left(P_{it}(Q_{1t}^{SHP_{Cn}}) - c_{it}^{SHP}\right).Q_{it}^{SHP_{Cn}} - \left(P_{it}(Q_{it}^{SHP} + Q_{it}^{CHPE_{Co}}) - c_{it}^{SHP}\right).Q_{it}^{SHP_{Cn}} + \theta.\left(Q_{it}^{SHP_{Cn}2} - (1 - \varphi_{it}).Q_{it}^{SHP_{Cn}2}\right)\right] - \Delta\pi_{it}.\left(\frac{z^{SHP}}{z^{SHP} + z^{CHPE} + z^{HUB}}\right) \tag{22}$$

$$\omega'_{it} = (1-\delta_t) \cdot \left[ \left( P_{it}(Q_{it}^{CHPE_{Cn}}) - c_{it}^{CHPE} \right) \cdot Q_{it}^{CHPE_{Cn}} - \left( P_{it}(Q_{it}^{SHP_{Cn}} + Q_{it}^{CHPE_{Cn}}) - c_{it}^{CHPE} \right) \cdot Q_{it}^{CHPE_{Cn}} \right] + \gamma \cdot \left[ (lt - y_t^{C_t} - \tau)^2 - (lt - y_t^{C_n} - \tau)^2 \right]$$

$$+ \rho \cdot \left[ y_t^{C_t} \cdot \log(Q_{it}^{CHPE_{Cn}} + 1) - (1 - \varphi'_{it}) \cdot y_t^{C_n} \cdot \log(Q_{it}^{CHPE_{Cn}} + 1) \right] - \Delta\pi_{it} \cdot \left( \frac{z^{CHPE}}{z^{SHP} + z^{CHPE} + z^{HUB}} \right) \tag{23}$$

## 4. Numerical study

This section presents an analysis of the behavior of the integrated system described in Fig. 1 and its performance. The analysis uses a case study and numerical examples to investigate the three game-theoretic models described in Section 3.

### 4.1. Data gathering for the hydrogen production in Atlantic Canada

Atlantic Canada is poised to develop low-carbon hydrogen projects, capitalizing on its wind and renewable energy resources. Various small-scale initiatives, such as the World Energy GH2 project in Newfoundland and Labrador, the Port of Belledune hydrogen plant in New Brunswick, and the EverWind Fuels project in Nova Scotia, focus on green hydrogen production via electrolysis [36]. These projects collectively aim for a capacity of 36,000 tons annually, potentially reaching 135,000 tons by 2035. Additionally, plans for blue hydrogen production, with a potential annual output of 165,000 tons, are underway at the Port of Belledune and Irving refineries. The marginal cost of hydrogen production in these projects is approximately $0.0002 per cubic ton. Delivery times average around two days ($lt^{SHP} = 2$), influenced by the proximity of production facilities to distribution hubs.

CHPE in Atlantic Canada is trying to gain a share of the growing (green) hydrogen market. The Atlantic Hydrogen Alliance is a consortium of organizations, including OERA, the Port of Halifax, Heritage Gas, Liberty Utilities, Saint John Energy, Atlantica Centre for Energy,

Deloitte, and Econext [37]. TransAlta Corporation and its subsidiary have been partnering with NB Power for over a decade to generate wind power, part of which is to produce green hydrogen. They operate the Kent Hills Wind Farm, New Brunswick's first wind farm, which has been in commercial operation since December 2008 [38].

Major hydrogen producers in Atlantic Canada produce about 500,000 tons of hydrogen annually for industrial use, of which about 60 % is grey hydrogen, 38 % blue hydrogen, and 2 % green hydrogen. There is an effort by these producers and others for a 200 % increase in the production capacity of low-emission hydrogen over a decade after accounting for ongoing and planned projects. It takes these large-scale producers up to 5 days to deliver orders after customers place them. Considering a natural gas price of $3.5/gigajoule, the estimated cost of grey hydrogen would be about $1.54/kg. Blue hydrogen comes at a cost of about 40 % higher than grey hydrogen due to the additional costs of carbon capture and storage [39]. With a carbon price of $170/ton, blue hydrogen would be around $2.16/kg. As for green hydrogen, its cost varies from $3 to $8 per kilogram, depending on the source of renewable electricity [39]. However, there is an anticipation that the average cost of green hydrogen will decrease to $1.50/kg by 2030 [40]. This study relies on the forecasted global production cost, both average and optimal hydrogen costs [13]. Additional details and parameters used in the models are in Table G1 in Appendix G.

### 4.2. Lump-sum payments in the coordination scheme

Analyzing the three collaboration structures requires first denoting $\omega_{it}$ and $\omega'_{it}$ as the lump-sum payments charged to the CHPE and SHP, respectively. In this subsection, the feasible areas of the coordination parameters, as formulated in Proposition 5, are determined for two hydrogen types produced by both entities. The optimal values for the associated coefficients for green and blue hydrogen within the coordination scheme for each period were calculated using Eqs. (18) and (19). The analysis was conducted employing the projected average costs of hydrogen over the period from 2026 to 2035. Fig. 2 illustrates the portion of costs contributed by the hydrogen hub, where $t$ represents different years within this timeframe. Additionally, $\varphi_t$ and $\varphi'_t$ are the discount factors applied within the tariff contracts, reflecting the agreements between the hydrogen hub and SHP, and hub and CHPE for green and blue hydrogen, respectively. Within the parameter ranges depicted in the 3D plots of Fig. 2, the coordination setting is accepted by all members, leading to improved profits compared to those under the market-based pricing strategy. The size of a contractual lump-sum payment paid to the hub by the SHP, $\omega_{it}$, or the CHPE, $\omega'_{it}$, is bound. Payment values falling outside designated bounds hinder the implementation of collaborative tariff contracts that coordinate hydrogen pricing and production decisions, affecting the system's profitability and coordination efficiency.

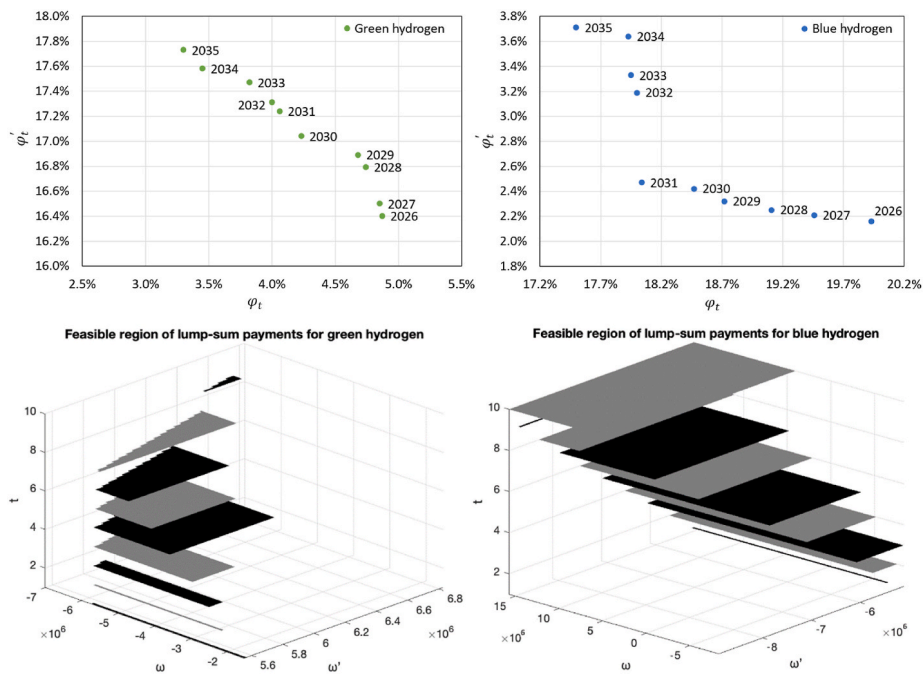

**Fig. 2.** Hydrogen hub's contribution to costs and feasible area for lump-sum payments.

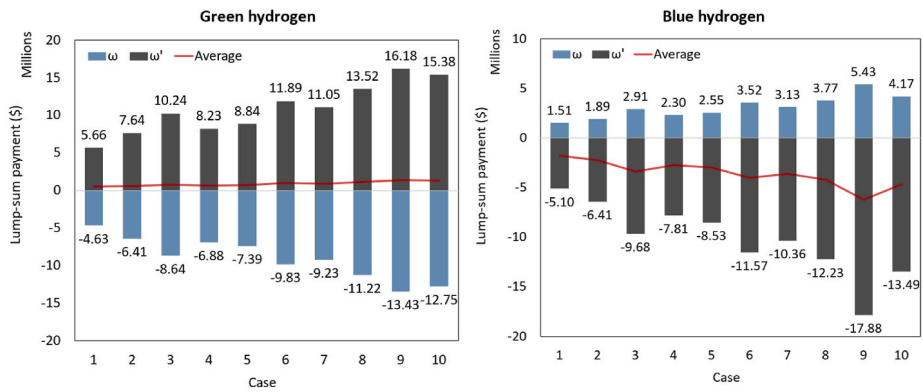

**Fig. 3.** Lump-sum payments imposed to members in the coordination scheme.

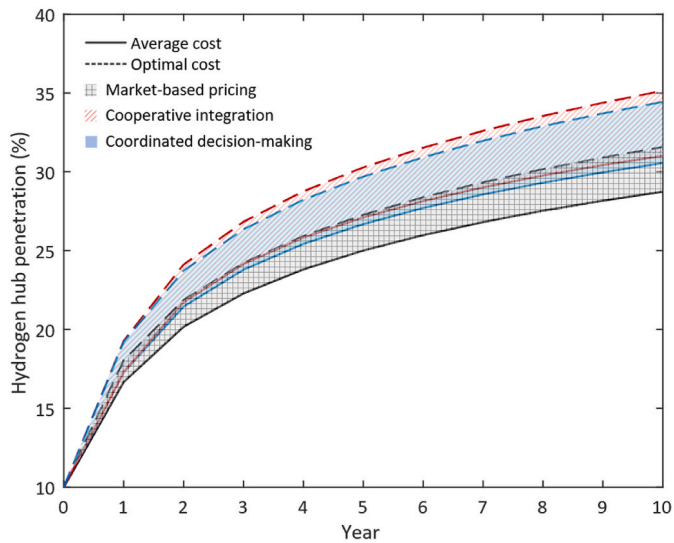

**Fig. 4.** Time path of hydrogen hub penetration.

The type of coordination contract and the profit-sharing mechanism determine the size of $\omega_{it}$ or $\omega'_{it}$, where surplus profits are distributed to participating members, each according to its bargaining power. Ten cases with varying bargaining power, detailed in Table G2 in Appendix G, were generated to examine how members receive a share of the surplus profit based on their bargaining strength. The bargaining power of the hub in these cases ranges from 0.1 to 0.325, as shown in Table G2. Fig. 3 shows the lump-sum payments imposed on hydrogen producers under the coordination tariff contracts for 2030, with the y-axis representing monetary values (e.g., millions of dollars). Notably, smaller-scale producers, relying more on RERs and contributing significantly to green hydrogen production, were charged by the hydrogen hub but received lump-sum payments for blue hydrogen. The trend in Fig. 3, where the CHPE receives payments to produce green hydrogen, suggests an opportunity for authorities to incentivize and promote producing that type of hydrogen. The non-linear trend in the figures results from the varying bargaining power and their effects on profit distribution. For instance, as the bargaining power of the hub ($z^{HUB}$) increases, hydrogen producers contribute more, resulting in lump-sum payments gradually increasing. Thus, decision-makers should strategically assess the varying bargaining powers of each member to foster a collaborative environment for sustainable growth in clean hydrogen production.

Throughout subsequent analyses, data derived from the first case was utilized for comprehensive examination and comparison.

### 4.3. Analyzing hydrogen hub penetration and market dynamics

This subsection aims to help understand how the hydrogen hub's penetration influences market share and profitability. Fig. 4 illustrates the evolving landscape of hydrogen hub penetration over time. Market penetration for the three coordination structures was assessed based on both average hydrogen production costs and optimal production costs projected by McKinsey & Company [13]. Across all three scenarios, the hub's penetration has consistently improved over time. In general, penetration increases in the order: Ct < Co ≈ Cn, which indicates that when decisions among hydrogen producers are made independently, as in the market-based pricing structure (Ct), the hub is less efficient compared to cooperative (Co) and coordinated (Cn) settings. The analysis reveals that the cooperative integration structure significantly influences penetration, increasing from 10 % to 35.14 %. However, under the coordination setting, penetration surpasses that of the market-based pricing structure, reaching 34.44 % in the next decade. The initial phase witnesses a steep rise in penetration, followed by a more gradual increase. This analysis projects that market penetration will grow further by 2035. This anticipated growth is attributed to a decrease in hydrogen prices with an increase in the proportion of green hydrogen. The findings underscore the dynamic nature of the hydrogen market, emphasizing its potential for expansion and evolution over the coming years.

The growing penetration of the hydrogen hub not only enhances profitability but also has a significant impact on market share. Fig. 5 illustrates the dynamic shifts in market share across various hydrogen types under three distinct scenarios based on average production costs. The analysis shows that while grey hydrogen initially dominated the market, constituting approximately two-thirds of the market share in the first year with a production of around 500,000 tonnes, its share witnessed a notable decline of about one-fourth over the subsequent decade. In contrast, the market share of green hydrogen tripled during this period. Notably, the cooperative integration strategy has emerged as the most effective approach for sustainable growth. Under this strategy, the share of green hydrogen production reached 19.06 %, generating 152,392 tonnes of hydrogen. This result underscores the viability and success of cooperative integration in helping the growth of environmentally conscious hydrogen production. Green hydrogen is expected to play a more substantial role in the clean energy transition in Atlantic Canada, followed by blue hydrogen. Grey hydrogen, while prevalent, may face challenges due to carbon emissions. Although green hydrogen may have a lower price with likely a growing preference for cleaner hydrogen by 2035, the region's green hydrogen production capacity is expected to remain below that of grey hydrogen.

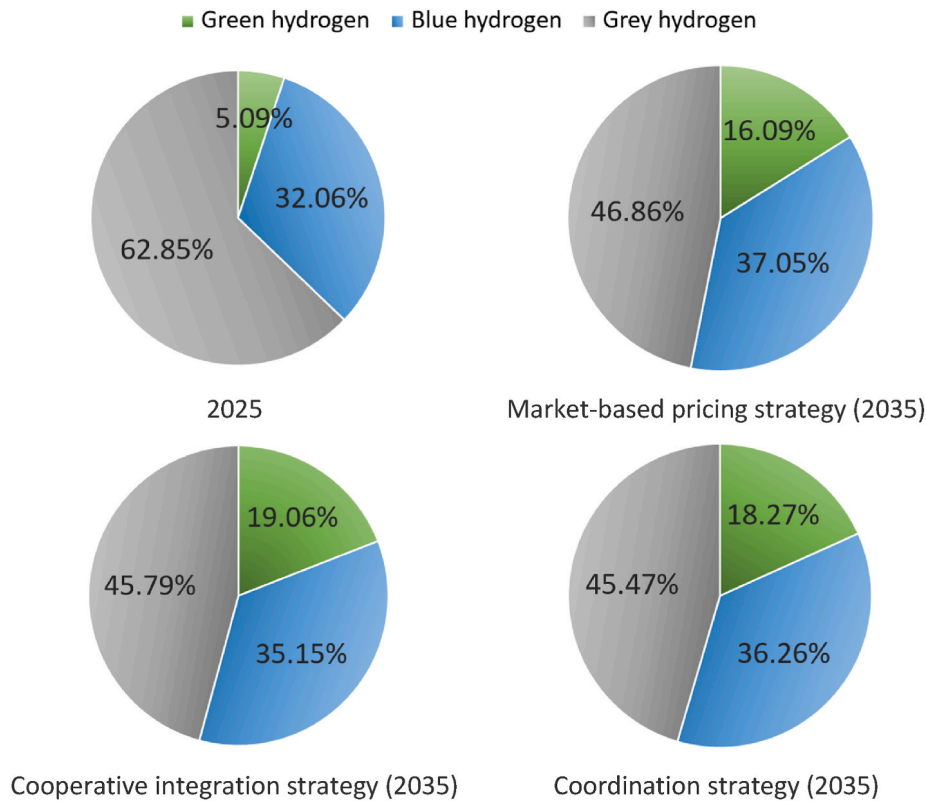

**Fig. 5.** Market share variations among different hydrogen types in the region.

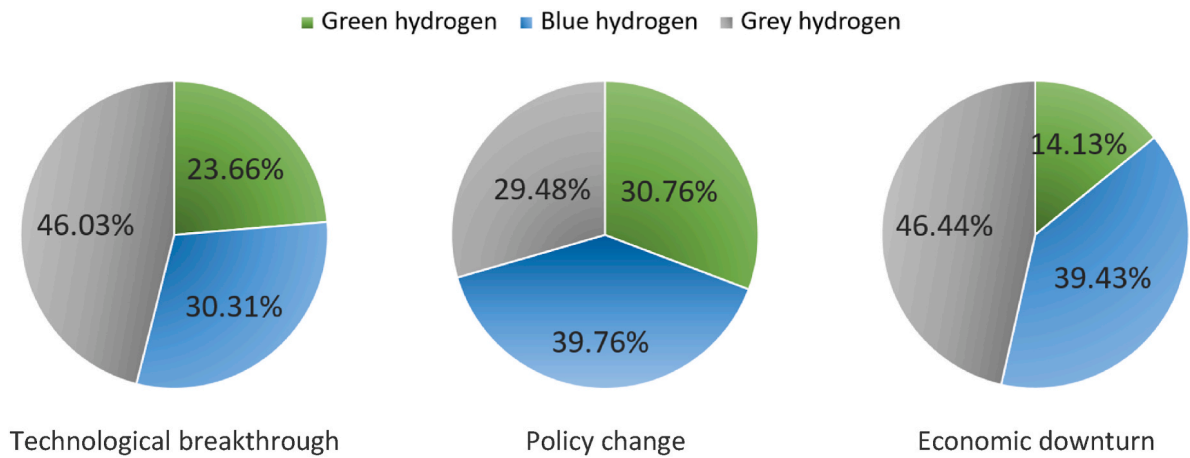

**Fig. 6.** Market share variations in 2035 under three exogenous shocks.

Given that grey hydrogen's market share is shrinking, managers should explore opportunities to diversify hydrogen production towards more environmentally friendly options. Prioritizing the production of green hydrogen aligns with market trends and enhances competitiveness. The analysis indicates that the decentralized system, as approached by the market-based pricing strategy, represents an efficient choice for decision-makers with environmental concerns. This approach supports sustainable growth and contributes to the increasing market share of green hydrogen over time. Additionally, decision-makers should explore investments in technologies and processes that facilitate the efficient production of green hydrogen. This approach positions companies to capitalize on the growing demand for green hydrogen.

Hydrogen market projections could shift dramatically with unexpected shocks. Current hydrogen market trends suggest analyzing three potential scenarios to estimate market share by 2035 under different strategies. <u>Scenario 1</u>: Technological breakthrough in green hydrogen production: For this analysis, the marginal cost of hydrogen production

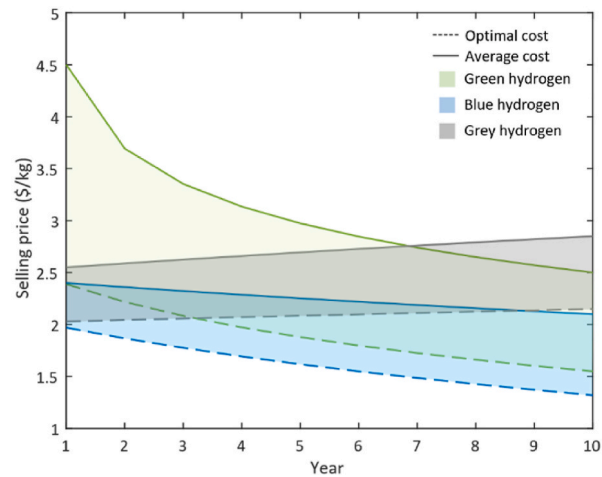

a)  Equilibrium price in a Cournot game

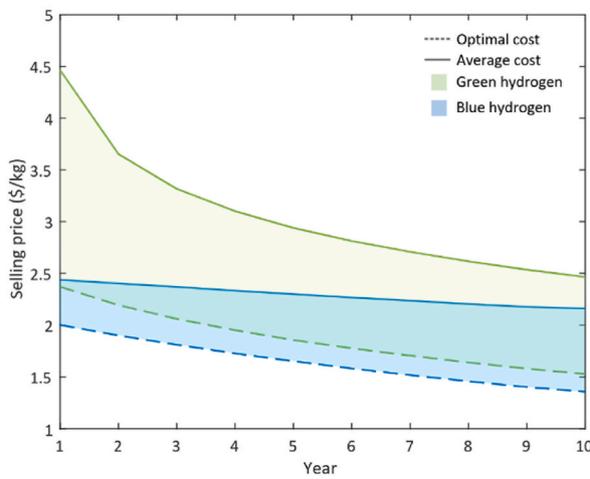

b)  Optimal price offered by SHP in Collusion game

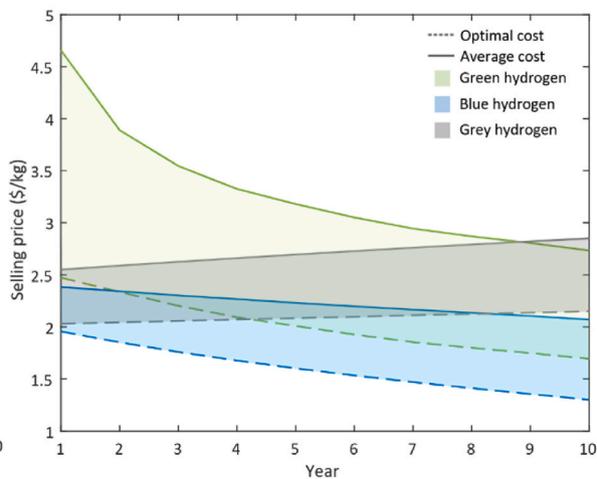

c)  Optimal price offered by CHPE in Collusion game

**Fig. 7.** Optimal price under market-based pricing and cooperative integration structures.

was reduced from $0.0002 per cubic ton to $0.00012 per cubic ton as technology became cheaper with each advancement. The demand sensitivity of green hydrogen prices was lowered from 0.2 to 0.15, indicating greater competitiveness and lower price sensitivity. The market penetration sensitivity for green hydrogen ($\alpha_t$) was increased from [0.1, 0.7] to [0.4, 0.8] to account for the perceived environmental benefits of technological breakthroughs. Scenario 2: Major policy change favoring clean energy: This analysis adjusted the cross-elasticity effect of the average price of green and blue hydrogen from [0.001, 0.009] to [0.004, 0.012], reflecting faster demand responses to favorable policy changes. The time sensitivity of green hydrogen, $s_2^i$, was reduced from 0.05 to 0.03, highlighting more efficient production and distribution enabled by policy support. The initial penetration level was also increased from 10 % to 15 % to represent accelerated market growth due to favorable policies. Scenario 3: Economic downturn: To assess this scenario, the retail price of gasoline and diesel, $q_t$, was increased from $2.25/kg to $2.625/kg, reflecting higher costs during an economic downturn, which affects fuel alternatives. The overall market demand factor was in the range [2,000, 3900], which captures reduced demand in a weakened economy. The substitution effect was increased from 0.05 to 0.07 to reflect stronger price sensitivity as consumers and businesses opt for cheaper alternatives.

The results, as shown in Fig. 6, illustrate how market shares might shift in 2035 under these different scenarios. In Scenario 1 (technological breakthrough), the market share of green hydrogen is expected to increase to 23.7 % due to reduced costs and higher efficiency. Blue hydrogen, however, slightly decreases to 30 % as green hydrogen becomes more competitive. Grey hydrogen further declines to 46 %, losing ground to cleaner alternatives. In Scenario 2 (policy change), green hydrogen's market share substantially increases to 31 %, driven by subsidies and stricter carbon regulations, while blue hydrogen moderately rises to 39.8 % as a transitional solution. Grey hydrogen's share significantly declines to 29.5 % due to regulatory pressures and carbon pricing. In Scenario 3 (economic downturn), green hydrogen experienced slower growth, reaching around 14 % due to higher initial costs. However, blue hydrogen remains stable or increases slightly to around 39 %, while grey hydrogen stabilizes or increases slowly to 46 % due to its lower production costs. These results underscore the importance of adapting strategies to changing market conditions and highlight the potential for green hydrogen to gain significant market share, particularly with technological advancements and supportive policies.

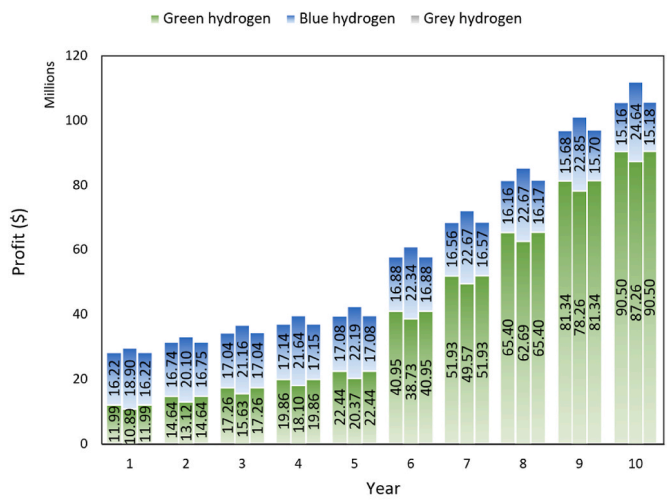

a) Profit dynamics of the SHP

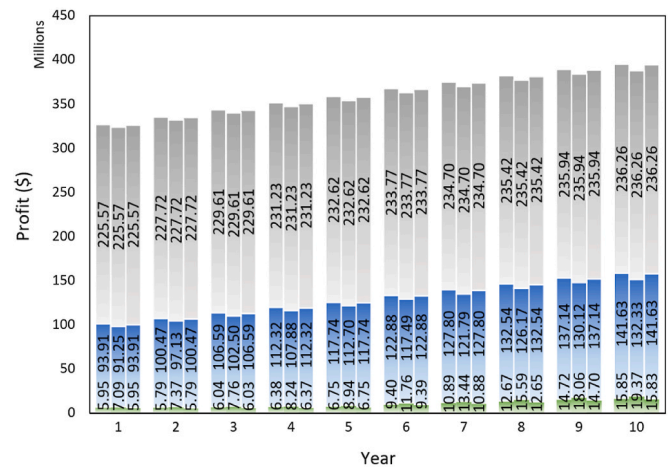

b) Profit dynamics of the CHPE

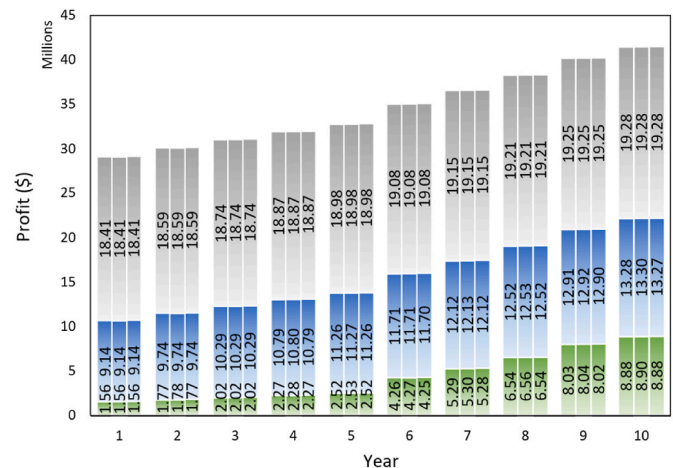

c) Profit dynamics of the hydrogen hub

**Fig. 8.** Members' profit for three hydrogen types under different models, namely Ct (first bar), Cn (second bar), and Co (third bar).

### 4.4. Profitability over the demand sensitivity to the price

This subsection examines the profitability of the demand sensitivity to price variations. Evaluating collaborative strategies and gaining insights into the future of hydrogen should involve comparing prices and profits under different structures. The prices determined through collusion methods are used in the coordinated decision-making structures (third method). Thus, this investigation explores optimal pricing within two other frameworks, i.e., market-based pricing and cooperative integration structures. The results for three distinct hydrogen types are shown in Fig. 7. Under the coordination of pricing by all hydrogen market producers, with both CHPE and SHP adopting the same pricing approach, green hydrogen experiences the most significant price decrease, as depicted in Fig. 7(a), plummeting by about 44 % (from $4.5/kg to $2.5/kg) in 10 years. Blue hydrogen, however, encountered a more modest decline of 12.5 %, reducing from $2.4/kg to $2.1/kg. Grey hydrogen sees an 11 % increase in price (from $1.8/kg to $2.8/kg; these values include a carbon tax). These findings align with what is reported in McKinsey & Company [13]. In the collaborative model presented in Fig. 7(b), where hydrogen producers align their efforts to maximize total profit, the optimal prices exhibit slight variations based on the market share of each producer. This observation could be because fossil fuel costs will increase in the next decade and beyond, thus shifting consumer preference towards green energy.

In the collaborative model presented in Fig. 7(b), where hydrogen producers align their efforts to maximize total profit, the optimal prices exhibit slight variations based on the market share of each producer. Notably, for green hydrogen, produced predominantly by SHP, the optimal prices are consistently lower than those offered by CHPE. For instance, the price for green hydrogen from SHP starts at $4.46/kg in the first period and decreases to $2.47/kg in the last period, whereas CHPE offers prices ranging from $4.65 to $2.73/kg over the same period. For blue hydrogen, the dynamics are reversed, with optimal prices offered by CHPE consistently lower than those from SHP. Over ten years, the price for blue hydrogen from CHPE dropped from $2.38 to $2.07/kg, while SHP offered slightly higher prices ranging from $2.44 to $2.16/kg. It is worth noting that grey hydrogen, produced by a single player in Atlantic Canada, remains unaffected by collaboration models. These findings emphasize the impact of cooperation or individual operation on hydrogen pricing and profitability and encourage collaboration among hydrogen producers by balancing economic viability and environmental sustainability. Additionally, incentivizing research and development in green hydrogen production technologies may contribute to further price reductions and promote a more sustainable energy landscape. Variations in confidence intervals for optimal pricing for both green and blue hydrogen are presented in Figure H1 in Appendix H.

Hydrogen price fluctuations significantly influence customer demand and producer profitability. As demonstrated in Fig. 5, the market shares of green, blue, and grey hydrogen shift over time under different strategies. These variations are critical when evaluating profitability, as they directly impact the profits shown in Fig. 8. Appendix I, Table 1, provides detailed profit data for each member, while Fig. 8 illustrates profitability across different structures. Market-based pricing (Ct) and cooperative integration (Cn) strategies do not consistently benefit all producers. Cooperative integration yields higher total profits, whereas market-based pricing favors members with larger market shares. Coordinated decision-making boosts SHP's green hydrogen profits but reduces CHPE's. Conversely, CHPE profits from blue hydrogen due to market dominance. The trends observed in Fig. 8 reflect the market share dynamics outlined in Fig. 8, with changes in market penetration influencing profit distribution. Projected cost reductions and production expansion in Atlantic Canada contribute to significant profit growth, albeit with slight reductions due to competition. While the hub incentivizes tariff contracts, grey hydrogen profits for CHPE remain unaffected. Coordinated contracts emerge as the only structure ensuring mutual benefits, suggesting a balanced approach for effective

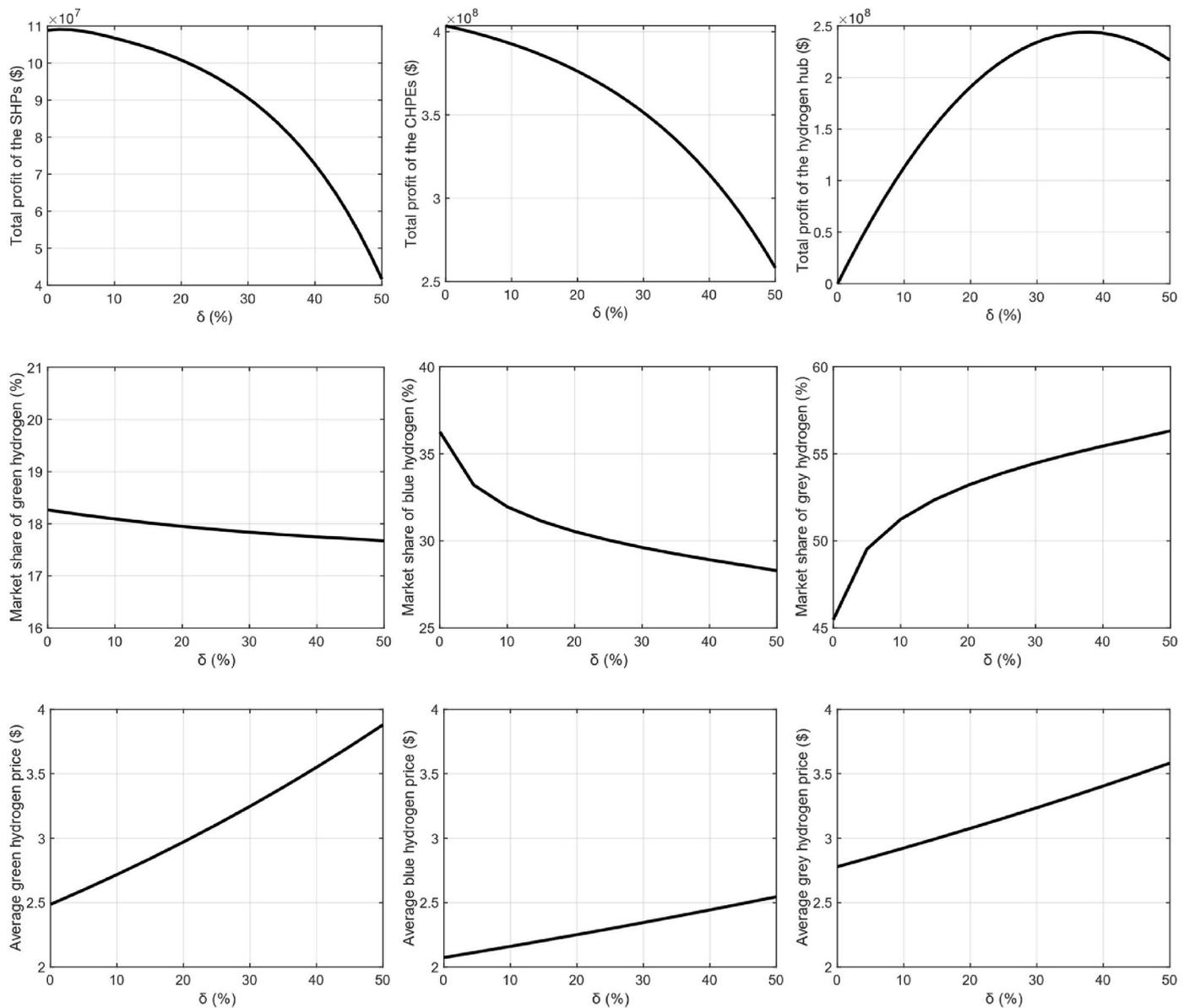

**Fig. 9.** Impact of commission fees on the offered price, market share, and total profit.

collaboration.

### 4.5. Influencing the hydrogen market: strategies for authorities and policymakers

Policies help provide the hydrogen industry with a roadmap that guides their investments, research, and development. They set the tone for sustainable growth and decarbonization. Well-defined policies balance the interests of various stakeholders, including producers, consumers, and environmental concerns. This subsection focuses on the impact of commissions imposed on both CHPE and SHP in terms of total profit, market share, and the offered price for each type of hydrogen, as depicted in Fig. 9. Specifically, the analysis involves varying the commission fee ($\delta_t$) from 0 to 50 % and assessing its effects on the coordinated decision-making structure in the last period (i.e., 2035).

The results from Fig. 9 reveal a negative impact on members' profits as the commission fee increases, subsequently influencing the pricing of different hydrogen types. The expected declining trend in prices for green and grey hydrogen, as observed in Fig. 7, is reversed, potentially leading to a loss in market share for these types. However, as analyzed in the coordinated decision-making structure, 0 % commission does not mean the absence of coordination under the hub. These fees are part of the coordinated tariff contracts aimed at ensuring mutual benefits and incentivizing clean hydrogen production. The incentives offered through participation in the hub go beyond immediate profit maximization, offering long-term strategic advantages, such as enhanced market access and collaborative opportunities. The analysis presented in Fig. 5 shows the share of grey hydrogen to decrease to 46 % in 2035; however, involving commission fee increment makes it surpass 56 %. Additionally, the findings indicate that commission or tax increments beyond 37 % fail to benefit the hydrogen hub or government. These results are due to significant increases in hydrogen prices, which slows demand and shrinks profits, as a result.

These results highlight the importance of effective policymaking to balance economic viability, environmental goals, and fair benefits for all stakeholders in the hydrogen market. Policymakers need to be careful not to impose many taxes that could upset how the market works and slow down the industry's progress. Designing the right policy is primary to encouraging new ideas, keeping the industry strong, and reaching environmental goals in the changing world of hydrogen.

## 5. Summary and conclusions

This study conducted a detailed analysis of the demand for 'Grey,' 'Blue,' and 'Green' hydrogen by considering key variables such as pricing, delivery lead time, market penetration, and collaboration strategies. The results indicated significant shifts in demand towards clean hydrogen, particularly 'Green' hydrogen, which is projected to see an average market share increase from 5.09 % in 2025 to 18.27 % by 2035 under coordinated strategies, representing a 258 % increase. Similarly, 'Blue' hydrogen is expected to grow from 32.06 % to 36.26 %, an approximate 13 % increase. These shifts highlight the need for policymakers, industry players, and investors to align with this evolving market. In terms of profitability, the study found that cooperative integration strategies can lead to a 10–15 % increase in total profits compared to market-based pricing for SHP. At the same time, CHPE sees a slight reduction in earnings from blue hydrogen under cooperative strategies. Coordinated strategies, on the other hand, increase profits for both SHP and CHPE by 8–12 %, demonstrating that collaboration between conventional and non-conventional hydrogen producers creates mutual benefits. Key parameters like commission fees were also analyzed in detail. For instance, increasing the commission fee from 0 % to 37 % reduced total profits by an average of 20 % across all hydrogen producers, emphasizing the need for a balanced policy approach. Additional sensitivity analyses revealed that a 10 % increase in production costs could decrease SHP's profitability by 15 %. Reducing delivery times by 10 % could boost demand and profits by 5–7%. These findings emphasize the importance of pricing strategies and environmentally friendly production methods in achieving a larger share of the hydrogen market. They also underscore the need for strategic pricing mechanisms and aligning policies with broader environmental goals.

### 5.1. Academic and practical contributions

This study makes several significant contributions to the academic literature on hydrogen market dynamics and collaboration strategies. It analyzed the effects of the factors influencing clean hydrogen demand, including pricing dynamics and market penetration effects, providing valuable industry insights. By evaluating three distinct collaboration models—market-based pricing, cooperative integration, and coordinated decision-making—the study elucidated the interplay between CHPE and SHP, enhancing the understanding of collaborative approaches in the hydrogen economy. Incorporating clean hydrogen rates and market penetration effects enriches the existing literature on hydrogen market modeling, offering a nuanced understanding of demand dynamics and strategic decision-making. This study bridged the gap between theoretical frameworks and practical applications in the hydrogen economy by integrating academic theories with empirical insights.

This study also provides valuable insights for policymakers, industry stakeholders, and investors navigating the evolving hydrogen market. The pricing dynamics and delivery lead time analysis offer actionable insights for hydrogen production and distribution companies, helping them develop strategic pricing mechanisms and optimize delivery systems to enhance market competitiveness. The evaluation of collaboration models reveals the potential for collaboration between CHPE and SHP, guiding the formation of strategic partnerships to address shared challenges in the hydrogen economy. Findings on market penetration effects and consumer preferences offer practical implications for companies anticipating demand shifts and tailoring market entry strategies. This study, in summary, inspires evidence-based decision-making, facilitating the transition toward a more sustainable and competitive hydrogen economy.

### 5.2. Managerial implications and insights

Hydrogen pricing is complex due to price-sensitive demand,

environmentally friendly production methods, and potential fuel substitution. Green hydrogen availability and the cross-elasticity effects of delivery lead time further complicate pricing. Decision-makers must balance consumer affordability with provider profitability. High prices may deter adoption, while low prices risk unsustainable business models. Strategic pricing can influence hydrogen demand and promote green hydrogen adoption through incentives. Timely and efficient operations justify premium pricing, with efficient logistics widening pricing margins. Dynamic pricing models reflecting real-time market changes help maintain competitiveness.

The findings emphasized implementing flexible policies that account for varying market conditions to ensure the overall resilience and sustainability of the hydrogen industry. Governments can incentivize collaboration and the transition to cleaner hydrogen types by offering tax discounts to producers engaged in coordinated integration. By providing tax incentives tied to environmentally friendly practices, authorities can encourage the adoption of cleaner technologies and foster a more sustainable hydrogen industry.

Authorities can organize the hydrogen market landscape by ensuring fair competition, preventing monopolies, and maintaining market stability. They may monitor hydrogen prices and the incentives they offer to produce clean hydrogen and manage price fluctuations. Promoting competition and diversification and providing subsidies, tax incentives, and grants, can support hydrogen producers, especially during transitions.

Early investments in hydrogen infrastructure can yield long-term benefits, informing strategic planning and resource allocation. Collaborative structures enhance market success through increased penetration, declining prices, and a growing share of green hydrogen. Understanding these dynamics helps decision-makers anticipate trends, plan for scalability, and optimize strategies for sustained profitability and relevance.

The proposed coordination between major conventional and small eco-friendly hydrogen producers fosters a balanced and environmentally conscious production ecosystem. This partnership supports a low-carbon economy with lower prices and economies of scale, making sustainable hydrogen competitive with fossil fuels. The tariff contract helps managers optimize resource allocation, reduce redundancy, and ensure a stable, environmentally compliant hydrogen supply.

To optimize the hydrogen market in Atlantic Canada, stakeholders should aim for a balanced market share, where green hydrogen's share grows to 20 % by 2035, while grey hydrogen declines to 45 %. Achieving this requires reducing green hydrogen production costs by 40 %, leveraging technological advancements and economies of scale, and targeting a price of $2/kg by 2035. As demonstrated in this study, coordination contracts present substantial benefits. For green hydrogen, it is recommended that in the first year, 2026, the lump-sum payment ($\omega_{it}$) to the SHP should range between $1.5 million and $4.6 million. Additionally, the lump-sum payments received from CHPE ($\omega'_{it}$) should be approximately $5.5 million. This coordination ensures mutual benefits, striking a balance between economic gains and environmental sustainability. When combined with a 15 % reduction in delivery times, hub penetration is projected to increase to 17 %. However, these outcomes are closely tied to production cost reductions. Policymakers can support this scenario by creating tax incentives linked to green hydrogen production and subsidizing technology development to reduce costs. Investors should prioritize funding projects that emphasize technological innovation and partnerships, while industry players should focus on collaborative strategies that increase market penetration and streamline logistics.

### 5.3. Limitations and future research directions

This study has some limitations. Firstly, the analysis relied on assumptions and simplifications inherent in modeling complex market dynamics, which may overlook real-world complexities. Secondly, the study focused on Atlantic Canada, limiting the generalizability of its findings. It excluded exporting hydrogen despite the international demand for it. Additionally, this study did not account for potential future regulatory or policy changes that could impact hydrogen market dynamics. While this study used predicted future costs of clean hydrogen, expected to fall below $2.5/kg by 2050, it is essential to acknowledge the significant challenges and uncertainties in achieving this target. Technological advancements, economies of scale, and policy support are examples of the factors that will help achieve cost reductions.

Future research should address these limitations to better understand hydrogen market dynamics. This study advocates for empirical studies to validate the model's findings and integrate factors such as policy interventions, technological advancements, and geopolitical developments. Investigating innovative collaboration models and investment strategies could provide valuable insights into maximizing the hydrogen economy's potential while addressing sustainability challenges. Comparative studies across different regions and industries can illuminate the variability of hydrogen market dynamics and the effectiveness of various strategies in diverse contexts. Continued research in this field is crucial for strategic decision-making and advancing the transition toward more sustainable energy.

## CRediT authorship contribution statement

**Mohammad Asghari:** Conceptualization, Methodology, Data curation, Visualization, Investigation, Validation, Writing – original draft, Writing – review & editing, Formal analysis, Resources, Software. **Hamid Afshari:** Conceptualization, Validation, Writing – review & editing, Supervision, Project administration, Funding acquisition. **Mohamad Y. Jaber:** Writing – review & editing, Supervision, Project administration, Funding acquisition. **Cory Searcy:** Writing – review & editing, Supervision, Project administration, Funding acquisition.

## Declaration of competing interest

The authors declare that they have no known competing financial interests or personal relationships that could have appeared to influence the work reported in this paper.

## Acknowledgements

The authors thank the Natural Sciences and Engineering Research Council of Canada (NSERC) for their financial support. The authors also express their gratitude to the editor and anonymous reviewers for their constructive comments throughout the review process.

## Appendix A. Supplementary data

Supplementary data to this article can be found online at https://doi.org/10.1016/j.rser.2024.115001.

## Data availability

Data will be made available on request.